\pdfoutput=1 
\documentclass{JINST}
\usepackage{upgreek}
\usepackage{cite}

\title{Separation of $e^+e^-\to e^+e^-$ and $e^+e^-\to\pi^+\pi^-$ events
using SND detector calorimeter.}
\author{M.N. Achasov$^{a,b}$, K.I. Beloborodov$^{a,b}$,
A. S. Kupich$^{a,b}$\thanks{Corresponding author.} \\
\llap{$^a$}Budker Institute of Nuclear Physics, Siberian Branch of the Russian
Academy of Science,\\
11 Lavrentyev, Novosibirsk 630090, Russia\\
\llap{$^b$}Novosibirsk State University,\\
Novosibirsk 630090, Russia\\
E-mail: \email{kupich@inp.nsk.su}}
   
\abstract{The technique of discrimination of the $e^+e^-\to e^+e^-$ and 
$e^+e^-\to \pi^+\pi^-$ events in energy range $0.5 < \sqrt{s} < 1$ GeV by
energy deposition in the calorimeter of SND detector was developed by applying
machine learning method. Identification efficiency for  $e^+e^-\to e^+e^-$ 
and $e^+e^-\to \pi^+\pi^-$ events in the range from 99.3 to 99.8 \% has been 
achived.}

\keywords{$e^+e^-$ annihilation, particle identification, calorimeter}

\begin{document}

\section{Introduction}

\begin{figure}[tbp]
\centering
\includegraphics[width=1.0\textwidth]{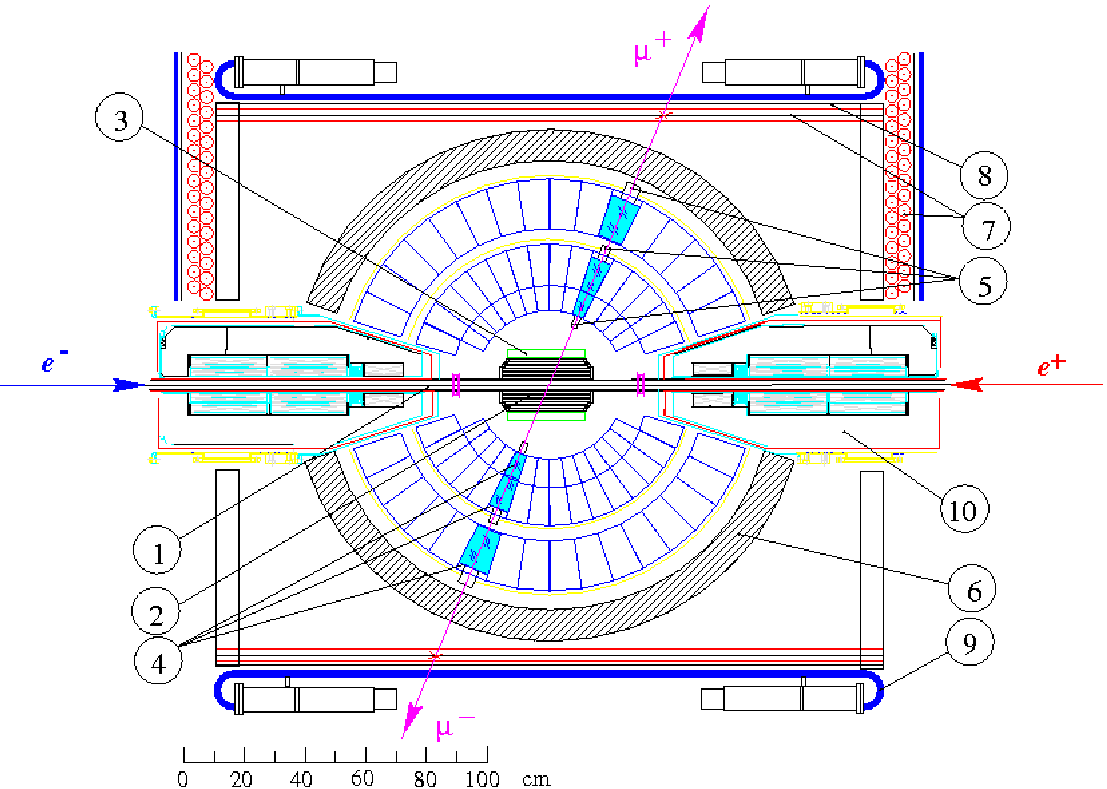}
\caption{SND detector, section along the beams: (1) beam pipe,
(2) tracking system, (3) aerogel Cherenkov counters, (4) NaI(Tl) crystals,
(5) vacuum phototriodes (6) iron absorber, (7) proportional tubes, (8) iron
plates, (9) scintillation counters, (10) solenoids of collider.}
\label{sndt}
\end{figure}

The spherical neutral detector SND \cite{snd} (Fig.\ref{sndt}) is a general 
purpose nonmagnetic detector operating at VEPP-2000 $e^+e^-$ collider in the
center-of-mass energy range from 0.2 to 2.0 GeV \cite{vepp2k}. Experimental 
studies include 
measurement of the cross sections of the $e^+e^-$ annihilation to hadrons. 
These measurements are  largely motivated by the need of high-precision 
calculation of the hadronic contribution to the anomalous magnetic moment of 
the muon $(g-2)/2$ \cite{g-2}. In particular, the $e^+e^-\to\pi^+\pi^-$ cross 
section at the energy region below 1 GeV gives the dominant contribution to
this value and should be measured with accuracy higher than 1\% \cite{cs2p}.

The cross section of the $e^+e^-\to\pi^+\pi^-$ process is measured in the
following way.
The collinear events $e^+e^-\to e^+e^-,\pi^+\pi^-,\mu^+\mu^-$ are selected.
The selected events are divided into two classes: $e^+e^-$ and
$\pi^+\pi^-,\mu^+\mu^-$.  The events of the $e^+e^-\to\mu^+\mu^-$ process are
subtracted according to the theoretical cross section, integrated  luminosity 
and detection efficiency.
The cross section of the $e^+e^-\to\pi^+\pi^-$ process is obtained as follows
\cite{cs}.
\begin{equation}\label{sde}
\sigma_{\pi\pi} = \frac{N_{\pi\pi}}{N_{ee}}
\frac{\varepsilon_{ee}}{\varepsilon_{\pi\pi}}
\frac{\sigma_{ee}}{1+\delta_r}
\label{sigpipi}
\end{equation}
Here $1+\delta_r$ is radiative correction, $N_{\pi\pi,ee}$ and 
$\varepsilon_{\pi\pi,ee}$ are the events numbers and detection efficiencies of
the processes $e^+e^-\to\pi^+\pi^-$ and $e^+e^-$ respectively, $\sigma_{ee}$
is cross section of the $e^+e^-\to e^+e^-$ process.

The $e^+e^-\to e^+e^-$, $\mu^+\mu^-$ and $\pi^+\pi^-$ events differ
by the energy deposition in the calorimeter. In  $e^+e^-\to e^+e^-$
events the electrons produce the electromagnetic shower with the most
probable energy losses of about 92\% of the initial particle energy.
Muons lose their energy by ionization of the calorimeter material through 
which they pass. The similar ionization losses as well as nuclear
interactions with the calorimeter material are experienced by charged
pions. The separation parameter of  $e^+e^-\to e^+e^-$ and 
$e^+e^-\to \pi^+\pi^-$ events with the energy $\sqrt{s}=$ 0.5 -- 1.0 GeV based
on these differences has been developed.

\section{The SND calorimeter}

SND detector \cite{snd} consists of the tracking system based on cilindrical 
drift and proportional chambers placed in a common gas volume, aerogel 
threshold counters \cite{ashif}, calorimeter and muon system based on 
proportional tubes and plastic scintillator. The solid angle of the tracking 
system is 94\% of $4\pi$ and the resolutions in the azimuth and polar angles 
0.45$^\circ$ and 0.8$^\circ$, respectively. The threshold Cherenkov counters 
are based on aerogel with refractive index 1.05. The  threshold momentums for 
$e/\mu/\pi$ are approximately equal to 1.6~/~330~/~436 MeV/c, respectively.
The solid angle of the system is about 60\% of $4\pi$.

\begin{figure}[tbp]
\centering
\includegraphics[width=1.0\textwidth]{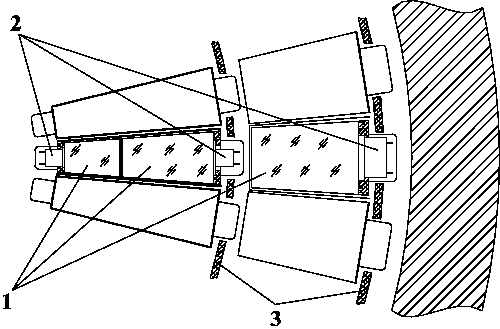}
\caption{NaI(Tl) crystals layout inside the calorimeter: (1) NaI(Tl) crystals,
(2) vacuum phototriodes, (3) aluminum supporting hemispheres.}
\label{cryst}
\end{figure}

The main part of SND is three-layer spherical electromagnetic calorimeter 
based on NaI(Tl) crystals \cite{snd}. Pairs of counters of the two
inner layers with thickness of 2.9 and 4.8 $X_0$ ($X_0=2.6$~cm) are sealed in 
thin (0.1 mm) aluminum containers, fixed to an aluminum supporting hemisphere
(Fig.~\ref{cryst}). Behind it, the third layer of NaI(Tl) crystals, 5.7 $X_0$
thick, is placed.  The total calorimeter thickness for particles 
originating from the interaction region is 34.7 cm (13.4 $X_0$) of NaI(Tl).
The total number of counters is 1632, the number of crystals per layer varies
from 520 to 560. The angular dimensions of the most of crystals are 
$\Delta\phi = \Delta\theta=9^\circ$, the total solid angle is $90\%$ of $4\pi$.

The scintillation light signals from the crystals are detected by vacuum 
phototriodes with an average photocathode quantum efficiency of about $15\%$ 
and the mean tube gain of about 10. The electronics of the calorimeter
consists of the charge sensitive preamplifiers with a conversion
coefficient of~0.7~V/pC, shaping amplifiers and 12-bit analog to digital
converter with a maximum input signal $U_{max}=2$ V. 

\section{Separation parameter}

The discrimination between electrons and pions in the SND calorimeter is
based on the difference in the energy deposition profiles for these particles. 
The energy depositions in the layers of the calorimeter towers, that form
the cluster  related to the particle are used. Here the tower is the three 
counters of the 1, 2 and 3 layers with the same $\theta$ and $\phi$ 
coordinates. In particular the following parameters are used: $^0E_{j}$ is the
energy deposition in  $j$th layer of the tower with the maximal energy 
deposition, $^1E_{j}$ is the sum of energy
depositions in $j$th layer of eight towers that  surround the tower with
the maximal energy deposition, $^2E_{j}$ is the sum of energy 
depositions in $j$th layer of the other towers of the cluster ($j=1,2,3$). 

\begin{figure}[tbp]
\centering
\includegraphics[width=1.0\textwidth]{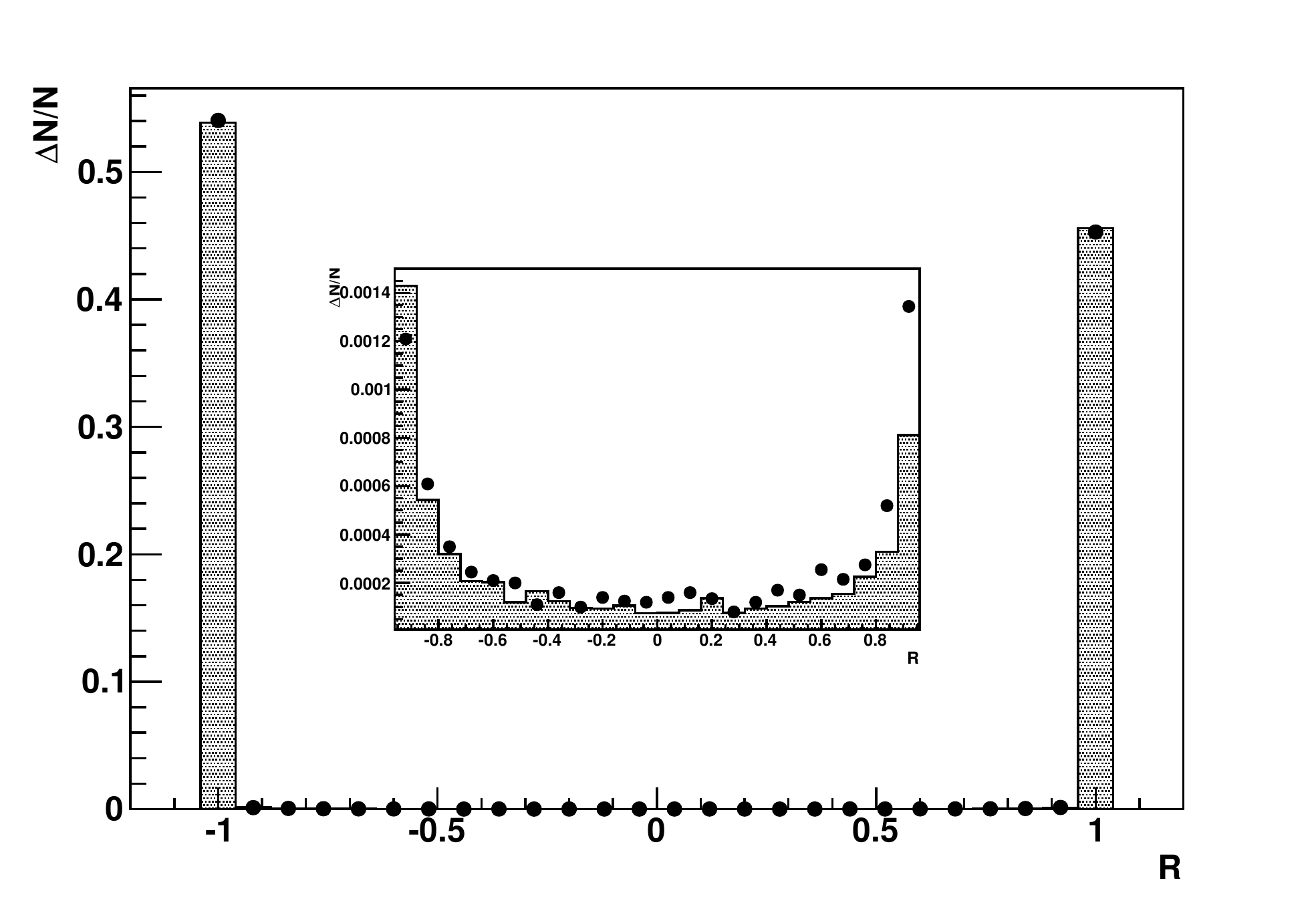}
\caption{The $e/\pi$ discrimination parameter $R$ distribution for all
collinear events at the energy $\sqrt{s}=778$ MeV. Dots -- experiment, 
histogram -- simulation.}
\label{bdt}
\end{figure}

In order to use the correlations between energy depositions in the calorimeter
layers in the most complete way, the corresponding separation parameter $R$ 
was based on the machine learning approach. For each energy point the boosted 
decision trees network (forest) has been constructed \cite{gbdt}. The forest 
includes 900 trees, the maximal depth of a tree is 9. The 18 energy 
depositions $^kE_j$  and the average polar angle 
$\theta_0=(\theta_1-\theta_2-180^\circ)/2$ of the particles have been used
as the discrimination variables. 
Here subscripts 1 and 2 denote the numbers of the particles. The training 
ensemble consists of simulated $e^+e^-\to\pi^+\pi^-$ and $e^+e^-\to e^+e^-$ 
events, that have passed the following cuts.
\begin{enumerate}
\item
$N_{cha}=2$. The events can contain neutral particles due to  nuclear
interactions of charged pions with detector material or due to electromagnetic
showers splitting.
\item
$|\Delta\theta|=|180^\circ-(\theta_1+\theta_2)|<8^\circ$ and
$|\Delta\phi|=|180^\circ-|\phi_1-\phi_2||<4^\circ$, where $\phi$ is the
particle azimuthal angle.
\item
$E_{1,2}>40$ MeV, where $E_i$ is the $i$th particle ($i=1,2$) energy 
deposition.
\item
$50^\circ<\theta_0<130^\circ$.
\item
The muon system $veto=0$.
\end{enumerate}
The output signal of the trained network (separation parameter) $R$ is a value
in the interval from -1.0 to 1.0  (Fig.\ref{bdt}). The $e^+e^-\to e^+e^-$ 
events are located in the region $R<0$, while $e^+e^-\to\pi^+\pi^-,\mu^+\mu^-$
events in $R>0$. 

\section{Identification efficiency}
\begin{figure}[tbp]
\centering
\includegraphics[width=1.0\textwidth]{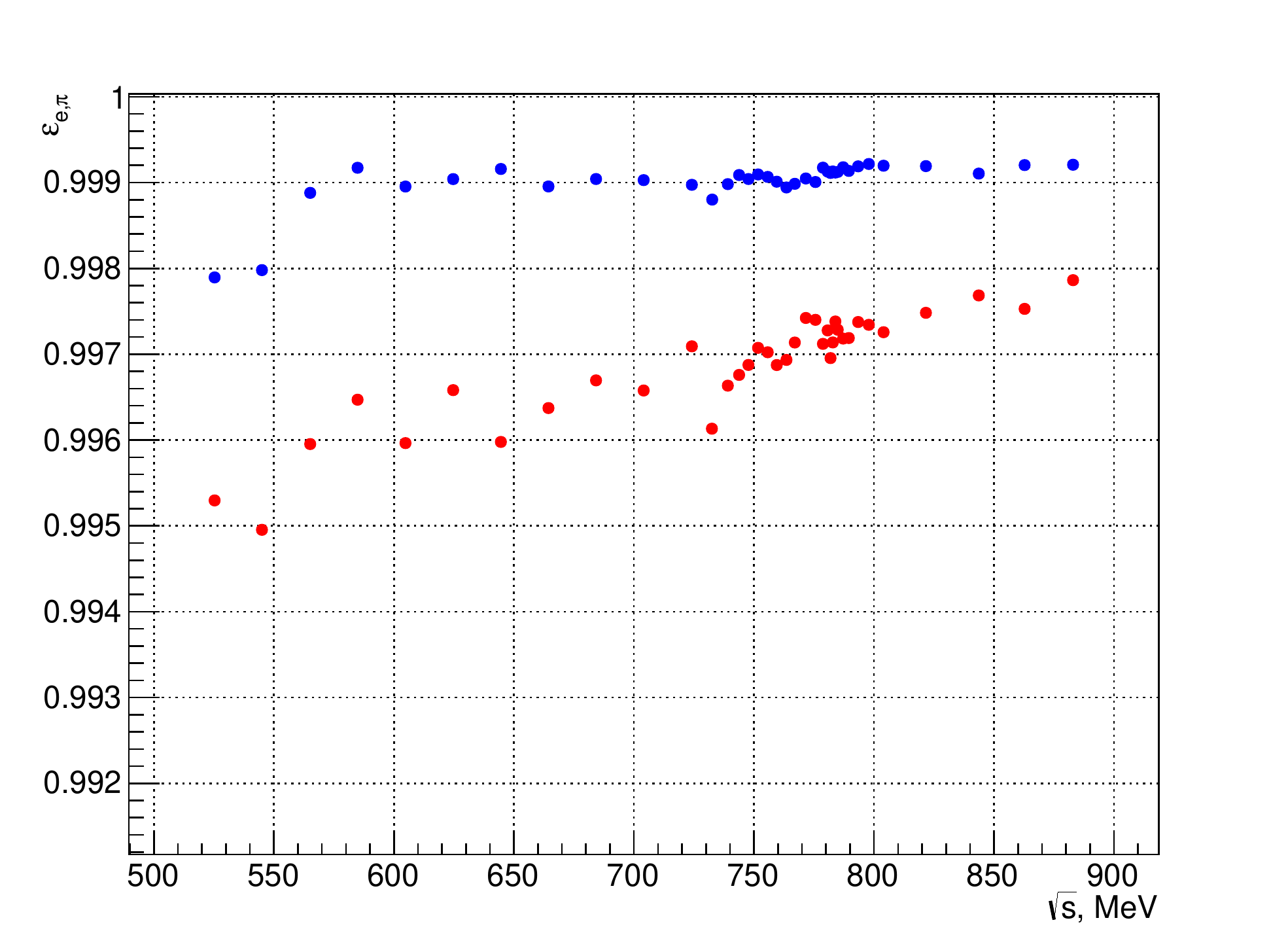}
\caption{Identification efficiencies of $e^+e^-\to e^+e^-$ (red dots) and
$e^+e^-\to\pi^+\pi^-$ (blue dots) events obtained using simulated events 
(the sample of $10^6$ events of each process in each energy point was used).
The statistical errors are less then the dots size.}
\label{ide}
\end{figure}

Identification efficiencies
\begin{eqnarray}\label{iden1}
\varepsilon_{e}=\frac{N^{ee}(R\in [-1;0])}{N^{ee}(R\in[-1;1])}, 
\end{eqnarray}
\begin{eqnarray}\label{iden2}
\varepsilon_\pi=\frac{N^{\pi\pi}(R\in ]0;1])}{N^{\pi\pi}(R\in [-1;1])}
\end{eqnarray}
of the processes $e^+e^-\to e^+e^-$ and  $\pi^+\pi^-$ obtained using simulated
events are shown in Fig.\ref{ide}. Here $N^{ee,\pi\pi}(R\in [a;b])$ are the 
numbers of events of $e^+e^-\to e^+e^-$  and  $\pi^+\pi^-$ processes in case 
if $R$ belongs to the interval $[a;b]$. The efficiencies (Fig.\ref{ide})
exhibit not a statistical spread from point to point. This can be explained by
the fact, that the number of the broken calorimeter channels are not 
coincident at different energy points.

Uncertainties in simulation of energy depositions in the calorimeter layers, in
particular, simulation of pions nuclear interactions, leads to an inaccuracy
in identification efficiencies. In order to estimate the systematic 
uncertainty of $e/\pi$ discrimination, the pseudo-$\pi\pi$ and pseudo-$ee$ 
events in the experiment and simulation have been formed in the  following way.

The pseudo-$ee$ event has been constructed from the particles of two separate 
collinear events demanding that their partners in these events look like
electrons (have $R^\prime<0$ and aerogel counter has been fired by both 
charged particles). Analogously, pseudo-$\pi\pi$ event has been constructed 
using events in which the aerogel counter hasn't been fired and $R^\prime>0$. 
Here $R^\prime$ is $e/\pi$ separation parameter based on the energy 
depositions $^kE_j$ of a single particle of the event. 
Identification efficiencies for simulated real and pseudoevents differes by 
0.02\% for $e^+e^-$ and 0.01\% for $\pi^+\pi^-$ events.

The experimental pseudo-$ee(\pi\pi)$ events contain small admixture of 
$\pi\pi(ee)$, $\mu\mu$, $e\pi$, $e\mu$, $\pi\mu$ events.
Due to this background,  identification efficiency for experimental
pseudo-$ee$ events is changed less then by $2\times 10^{-4}$ in the whole
energy region $\sqrt{s}=$ 0.5 -- 1.0 GeV.  In case of experimental
pseudo-$\pi\pi$ events, efficiency is changing less than by $2\times 10^{-4}$
for the energy $\sqrt{s}>0.6$ GeV, and below it changes up to
0.009 at the energy 0.5 GeV. The pseudo-$\pi\pi$ events 
with neglectable background contribution  and higher statistics for the low
energy region from 0.6 to 0.5 GeV have been constructed by using pions from 
the  $e^+e^-\to\pi^+\pi^-\pi^0$ reaction. In order to construct 
the pseudo $\pi\pi$ event with the pions with the energy $E_0$, two charged 
pions with energies $E_\pi$ such that $|E_0-E_\pi|<5$ MeV have been used from 
two separate $e^+e^-\to\pi^+\pi^-\pi^0$ events. The $e^+e^-\to\pi^+\pi^-\pi^0$
events collected at the peaks of $\omega$ and $\phi$ mesons were used. 
The pions energies $E_\pi$ have been obtained by using kinematic fit, which
was performed under the following constraints: the charged particles are 
assumed to be pions, the system has zero total momentum, the total energy is 
$\sqrt{s}$, and the photons originate from the $\pi^0\to\gamma\gamma$ decays.
The difference between identification efficiencies of simulated 
$e^+e^-\to\pi^+\pi^-$ and simulated pseudo-$\pi\pi$  events of this type is 
0.02\%. 

\begin{figure}[tbp]
\centering
\includegraphics[width=1.0\textwidth]{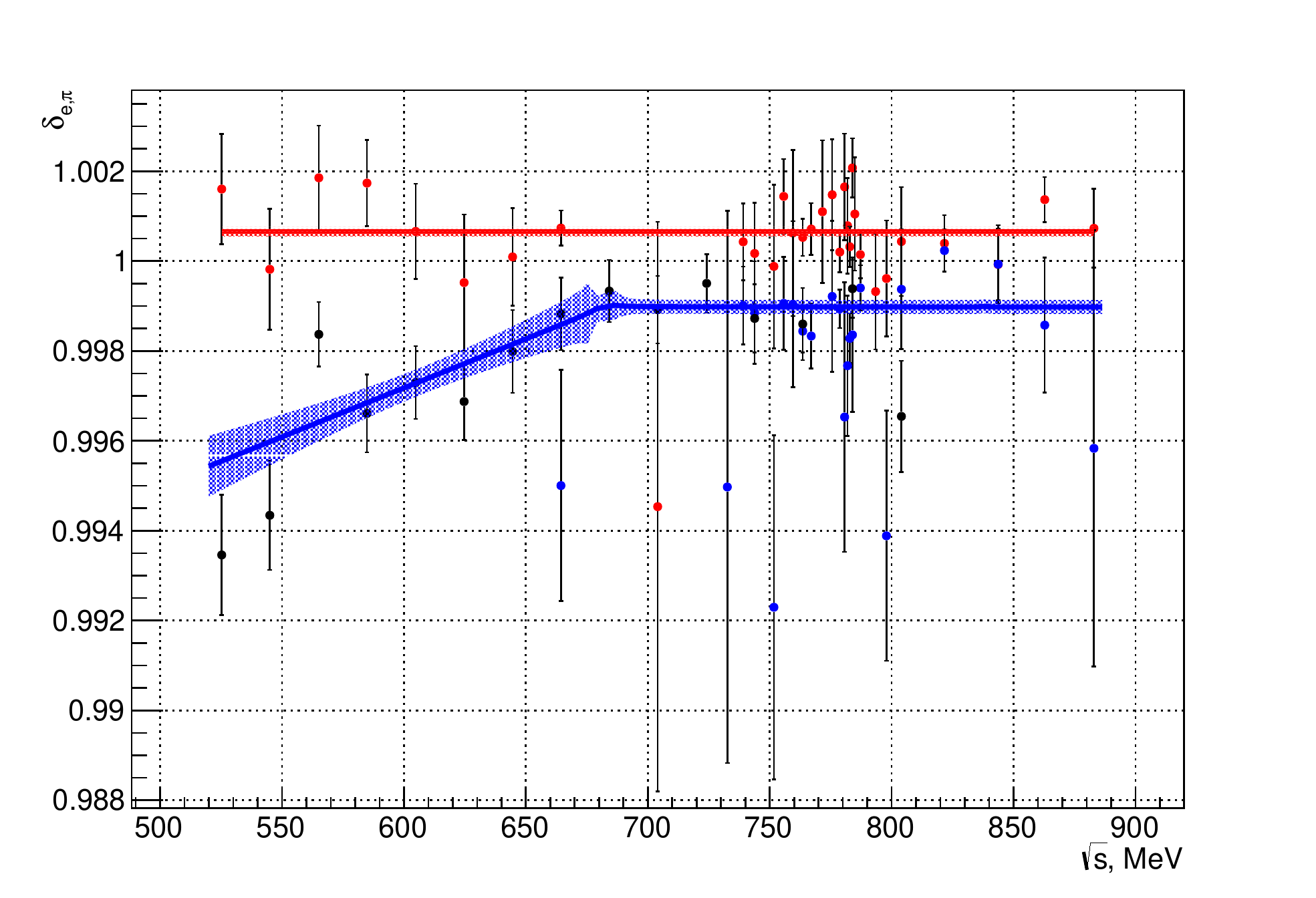}
\caption{Correction coefficients for identification efficiencies 
of $e^+e^-\to e^+e^-$ (red dots) and $e^+e^-\to\pi^+\pi^-$ (blue and black 
dots) events. The blue and black dots show the values of $\delta_\pi$ obtained
using pseudo $\pi\pi$ events constructed from $e^+e^-\to\pi^+\pi^-$ and 
$e^+e^-\to\omega,\phi\to\pi^+\pi^-\pi^0$ events, respectively. 
Lines are the results of approximations, dashed regions show the errors.}
\label{cor}
\end{figure}
Using pseudoevents the correction coefficients for identification
efficiencies of a real $e^+e^-\to e^+e^-$ and $e^+e^-\to\pi^+\pi^-$ events
has been obtained as follows
\begin{equation}
\delta_x=\frac{\epsilon^{exp}_x}{\epsilon^{mc}_x},
\end{equation}
where $x=e(\pi)$, $\epsilon^{exp}_x$ and $\epsilon^{mc}_x$ are identification
efficienties for experimental and simulated pseudoevents respectively. The
energy dependencies of the correction coefficients are shown in Fig.\ref{cor}.
The $\delta_e$ coefficient does not depend on energy, its average value is
equal to $1.0006 \pm 0.0001$. The values of correction coefficients 
$\delta_\pi$ obtained using pseudo $\pi\pi$ events constructed from 
$e^+e^-\to\pi^+\pi^-$ and $e^+e^-\to\omega,\phi\to\pi^+\pi^-\pi^0$ events are 
in agreement within their statistical errors. Their energy dependence was
fitted by the function
\begin{eqnarray}
\delta_\pi(\sqrt{s})=a\biggr(\sqrt{(\sqrt{s}-b)^2-10(\sqrt{s}-b)}-
(\sqrt{s}-b)\biggl)+c.
\end{eqnarray}
It was obtained that $\delta_\pi=0.9990\pm0.0002$ at the energy region
$\sqrt{s}$ above 0.65 GeV and below $\delta_\pi$ changes upto $0.9950\pm 0.0006$
at $\sqrt{s}=0.52$ GeV (Fig.\ref{cor}).

\begin{figure}[tbp]
\centering
\includegraphics[width=1.0\textwidth]{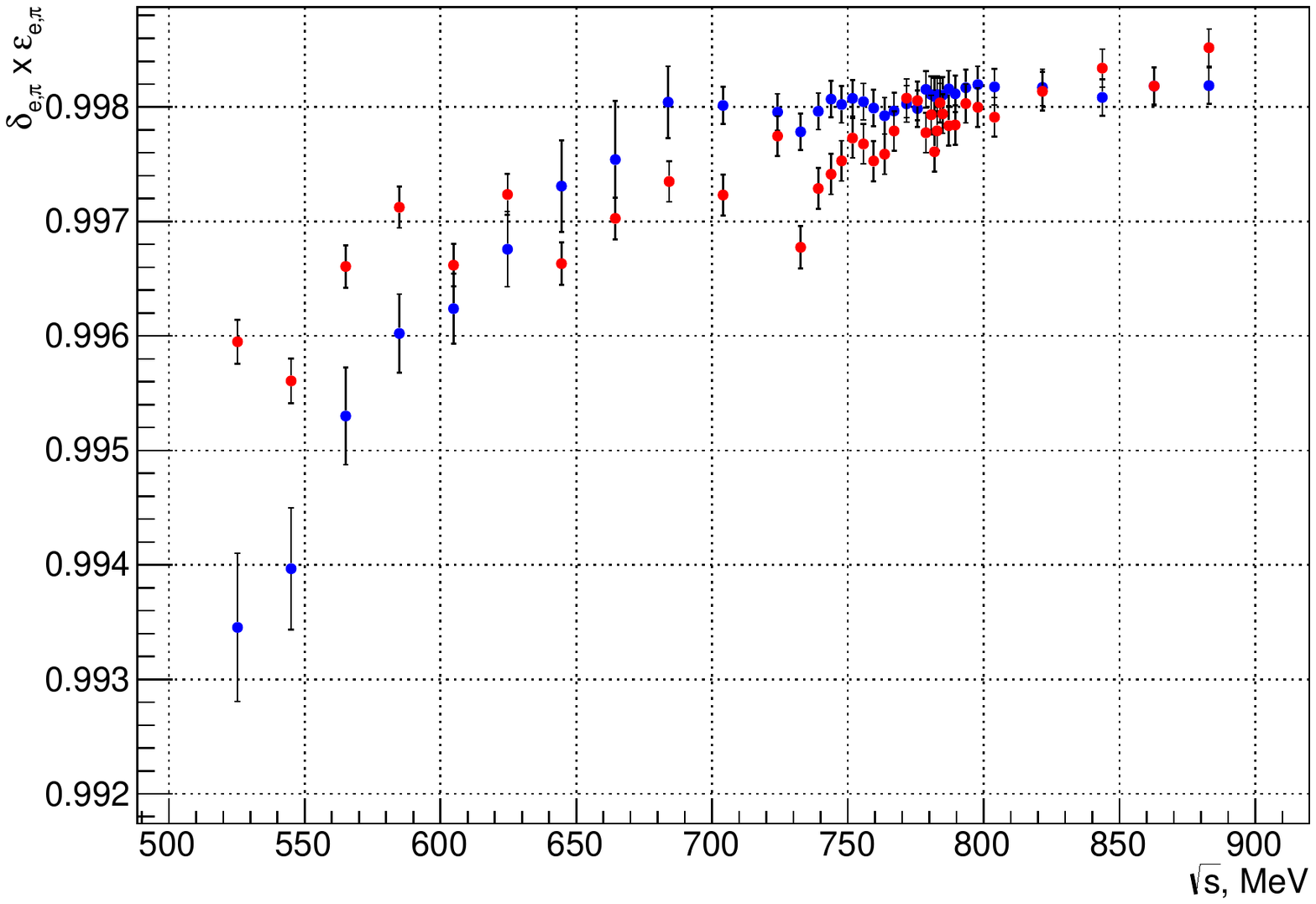}
\caption{Corrected identification efficiencies $\delta_e\varepsilon_e$ and
$\delta_\pi\varepsilon_\pi$ of $e^+e^-\to e^+e^-$ (red dots) and 
$e^+e^-\to\pi^+\pi^-$ (blue dots) events.}
\label{cide}
\end{figure}

\begin{table}[tbp]
\caption{Various contributions to the relative errors of the $\delta_e$ and 
$\delta_\pi$ correction coefficients.}
\label{tab1}
\smallskip
\centering
\begin{tabular}{|c|c|c|c|}
\hline
Error & Contribution to $\delta_e$, \% &
Contribution to $\delta_\pi$ &
Contribution to $\delta_\pi$ \\
 & & at $\sqrt{s}>0.65$ GeV, \% & at $\sqrt{s}<0.65$ GeV, \% \\
\hline
$\sigma_{stat}$ & 0.01 & 0.02 & 0.02 -- 0.06 \\
$\sigma_{ID}$   & 0.02 & 0.01 & 0.02 \\
$\sigma_{bkg}$  & 0.02 & 0.02 & -- \\
\hline
$\sigma_{tot}$  & 0.03 & 0.03 & 0.03 -- 0.06 \\
\hline
\end{tabular}
\end{table}

The total error of the correction coefficient determination is
\begin{equation}
\sigma_{tot} =\sigma_{stat}\oplus\sigma_{ID}\oplus\sigma_{bkg}.
\end{equation}
Here $\sigma_{stat}$ is the statistical error, $\sigma_{ID}$ is the
difference in identification efficiency for real and pseudoevents,
$\sigma_{bkg}$ is the change of identification efficiency for experimental
pseudoevents due to background admixture. The magnitudes for
various contributions to the total error are shown in table~\ref{tab1}.
The total relative error of $\delta_e$ is $\sigma_{tot}=0.03$\% and of 
$\delta_\pi$ is $\sigma_{tot}=0.03$\% at $\sqrt{s}>0.65$ GeV and 
$\sigma_{tot}=$ 0.03 -- 0.06 \% at $\sqrt{s}<0.65$ GeV.

\begin{figure}[tbp]
\centering
\includegraphics[width=1.0\textwidth]{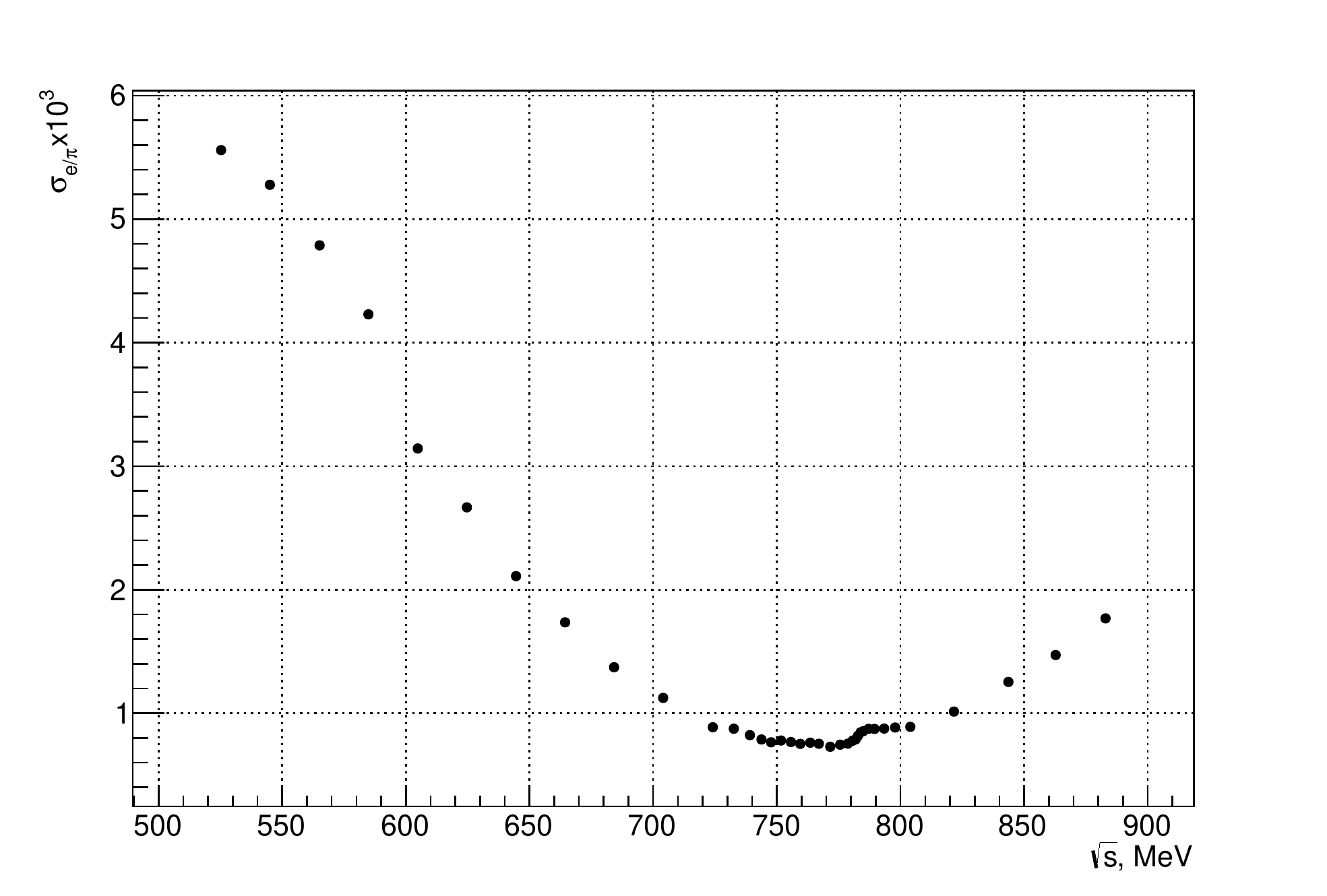}
\caption{The relative error $\sigma_{e/\pi}$ of $e^+e^-\to\pi^+\pi^-$ cross
section  due to the errors of identification efficiencies at the
different energy points.}
\label{sep}
\end{figure}

The corrected identification efficiencies of processes $e^+e^-\to e^+e^-$
and $e^+e^-\to\pi^+\pi^-$ are shown in Fig.\ref{cide}. Their errors are
dominated by the errors of the correction coefficients. Contribution of the
identification efficiencies errors to the total relative error of
$e^+e^-\to\pi^+\pi^-$ cross section (\ref{sigpipi}) is shown in Fig.\ref{sep}
and is less than 0.2\% for the most energy points.

\section{Conclusion}

The technique of discrimination of the $e^+e^-\to e^+e^-$ and 
$e^+e^-\to \pi^+\pi^-$ events using energy 
deposition in the calorimeter of SND detector has been developed. 
Identification efficiency for  $e^+e^-\to e^+e^-$ and $e^+e^-\to \pi^+\pi^-$ 
events has been obtained. Contribution of the identification efficiencies 
errors to the total error of $e^+e^-\to\pi^+\pi^-$ cross section is less than
0.2\% for the most energy points.

\acknowledgments

This work was supported in part  by the RFBR grants 14-02-00129-a and 
16-32-00542-mol-a, part of this work related to recunstruction of clusters in
calorimeter was supported by   Russian Science Foundation
(project N 14-50-00080).

\end{document}